\documentclass[numberedappendix,apj,twocolumn]{emulateapj}
\usepackage[bookmarks,bookmarksnumbered,colorlinks=true, citecolor=blue, linkcolor=black,breaklinks]{hyperref}
\usepackage{epstopdf}
\usepackage{apjfonts}
\usepackage{amsmath} 
\usepackage{threeparttable}
\usepackage{float}
\newcommand{\angstrom}{\textup{\AA}}

\newcommand{\fesc}{$f_\mathrm{esc}$}

\shorttitle{LyC Candidates at $z\sim2$ from the HDUV}
\shortauthors{Naidu et al.}

\submitted{Draft version \today}

\begin{document}

\title{The HDUV Survey: Six Lyman Continuum Emitter Candidates \\ at z$\sim$2 Revealed by HST UV Imaging\altaffilmark{1}}

\altaffiltext{1}{Based on observations made with the NASA/ESA Hubble Space Telescope, obtained from the data archive at the Space Telescope Science Institute. STScI is operated by the Association of Universities for Research in Astronomy, Inc. under NASA contract NAS 5-26555.}

\author{R. P. Naidu\altaffilmark{2,3},
P. A. Oesch\altaffilmark{4,3,5}, 
N. Reddy\altaffilmark{6,$\dagger$},
B. Holden\altaffilmark{7},
C. C. Steidel\altaffilmark{8},
M. Montes\altaffilmark{3},
H. Atek\altaffilmark{9},
R. J. Bouwens\altaffilmark{10}, \\
C. M. Carollo\altaffilmark{11}, 
A. Cibinel\altaffilmark{12},
G. D. Illingworth\altaffilmark{7}, 
I. Labb\'{e}\altaffilmark{10}, 
D. Magee\altaffilmark{7},
L. Morselli\altaffilmark{13},\\
E. J. Nelson\altaffilmark{14}, 
P. G. van Dokkum\altaffilmark{3,5}, 
S. Wilkins\altaffilmark{12}}

\altaffiltext{2}{Yale-NUS College, 12 College Avenue West, Singapore 138614}
\altaffiltext{3}{Astronomy Department, Yale University, New Haven, CT 06511, USA}
\altaffiltext{4}{Geneva Observatory, Universit\'e de Gen\`eve, Chemin des Maillettes 51, 1290 Versoix, Switzerland}
\altaffiltext{5}{Yale Center for Astronomy and Astrophysics, Yale University, New Haven, CT 06511, USA}
\altaffiltext{6}{University of California, Riverside, 900 University Ave, Riverside, CA 92507, USA}
\altaffiltext{7}{UCO/Lick Observatory, University of California, Santa Cruz, 1156 High St, Santa Cruz, CA 95064, USA}
\altaffiltext{8}{Cahill Center for Astronomy and Astrophysics, California Institute of Technology, MS 249-17, Pasadena, CA 91125, USA}
\altaffiltext{9}{Institut d'astrophysique de Paris, 98bis Boulevard Arago, 75014 Paris, France}
\altaffiltext{10}{Leiden Observatory, Leiden University, NL-2300 RA Leiden, The Netherlands}
\altaffiltext{11}{Institute for Astronomy, ETH Zurich, 8092 Zurich, Switzerland}
\altaffiltext{12}{Astronomy Centre, Department of Physics and Astronomy, University of Sussex, Brighton, BN1 9QH, UK}
\altaffiltext{13}{Excellence Cluster Universe, Boltzmannstr. 2, D-85748 Garching bei M\"{u}nchen, Germany}
\altaffiltext{14}{Max Planck Institute for Extraterrestrial Physics, Giessenbachstrasse, 85741 Garching bei M\"{u}nchen, Germany}
\altaffiltext{$\dagger$}{Alfred P. Sloan Foundation Fellow}

\email{rohan.naidu@u.yale-nus.edu.sg}

\begin{abstract}

We present six galaxies at $z\sim2$ that show evidence of Lyman continuum (LyC) emission based on the newly acquired UV imaging of the Hubble Deep UV legacy survey (HDUV) conducted with the WFC3/UVIS camera on the Hubble Space Telescope ($HST$). At the redshift of these sources, the HDUV F275W images partially probe the ionizing continuum. By exploiting the $HST$ multi-wavelength data available in the HDUV/GOODS fields, models of the UV spectral energy distributions, and detailed Monte Carlo simulations of the intergalactic medium absorption, we estimate the absolute ionizing photon escape fractions of these galaxies to be very high -- typically $>60\%$ ($>13\%$ for all sources at 90\% likelihood). Our findings are in broad agreement with previous studies that found only a small fraction of galaxies to show high escape fraction. These six galaxies comprise the largest sample yet of LyC leaking candidates at $z\sim2$ whose inferred LyC flux has been cleanly observed at $HST$ resolution. While three of our six candidates show evidence of hosting an active galactic nucleus (AGN), two of these are heavily obscured and their LyC emission appears to originate from star-forming regions rather than the central nucleus. This suggests an AGN-aided pathway for LyC escape from these sources. Extensive multi-wavelength data in the GOODS fields, especially the near-IR grism spectra from the 3D-\textit{HST} survey, enable us to study the candidates in detail and tentatively test some recently proposed indirect methods to probe LyC leakage--namely, the  [\ion{O}{3}]/[\ion{O}{2}] line ratio and the H$\beta-$UV slope diagram. High-resolution spectroscopic followup of our candidates will help constrain such indirect methods which are our only hope of studying $f_{esc}$ at $z\sim5-9$ in the fast-approaching era of the \textit{James Webb Space Telescope}.

\end{abstract}



\section{Introduction}

Identifying the sources that dominated cosmic reionization in the first 1 Gyr of cosmic time is still one of the key open questions of observational extragalactic cosmology. Recent advances in tracing the build-up of galaxies during the epoch of reionization (EoR) at $z>6$ indicate that ultra-faint galaxies are very abundant in the early universe and that they dominate the UV luminosity density. This has led several authors to speculate that the faint galaxy population is the main driver for reionization, a scenario which can reconcile several independent measurements of the reionization history \citep[e.g.,][]{Bouwens06a,Bouwens12b,Bouwens15c,Oesch09,Ouchi09,Bunker10, mclure10, Grazian12,Finkelstein12b, duncan&conselice15,Robertson15}.

The main unknown in these studies is the fraction of ionizing photons that escape galaxies into the inter-galactic medium (IGM), the so-called escape fraction, $f_\mathrm{esc}$. This remains unconstrained observationally during the EoR. The typical conclusion of reionization calculations is that the escape fraction of galaxies has to be $\gtrsim$10\%. Otherwise, their ionizing photon production falls short of the required value to complete reionization by galaxies, and other sources such as active galactic nuclei (AGN), are needed to contribute significantly \citep[e.g.,][]{Mitra15,Mitra16,Giallongo15,madau15,Price16,Feng16}. 

Direct observational constraints on \fesc\ are effectively impossible to obtain at $z\gtrsim4.5$ and into the EoR due to the high opacity of the intervening IGM absorption \citep[e.g.][]{Prochaska10,Inoue14}. However, at lower redshifts such direct studies of Lyman continuum (LyC) photons (at rest-wavelength $\lambda<912$ \AA) are possible.
 Until recently, the few constraints on \fesc\ that existed in the local universe were only upper limits indicating very low values of only a few percent at most \citep[e.g.][]{Leitherer95,Deharveng01,Grimes09, Siana10, Rutkowski16} --- far too small compared to the $\gtrsim10\%$ required for cosmic reionization.
However, recent work with the COS spectrograph on the \textit{HST} has identified a sub-sample of highly star-forming galaxies in the local universe that appear to show significant and detectable LyC emission \citep[][]{Borthakur14,Izotov16a,Izotov16b,Leitherer16}. 

At higher redshift, the situation is similar. At $z\sim2-3$, the LyC shifts into the observed $\sim 2000-3500~\angstrom$ range, allowing UV sensitive instruments to directly detect ionizing photons. Early observations resulted in confusing results with many of the direct detections being attributed to contamination by foreground sources \citep{Vanzella10, Vanzella12, Nestor11, Siana15, Mostardi15, Grazian16}. However, recently a small sample of three galaxies with confirmed direct detections of their LyC emission has emerged \citep{Mostardi15,Vanzella16, Shapley16}. 
One of these is $Ion2$ at $z=3.2$, which was originally identified in \citet[][henceforth V15]{Vanzella15} using a method similar to the one we adopt in this paper. In particular, V15 simulated the UV flux of sources with secure spectroscopic redshifts to identify candidate LyC sources in the GOODS-S broad-band imaging data. This resulted in two candidates, one of which, $Ion2$, has been confirmed as an LyC leaker through direct follow-up imaging \citep{Vanzella16, deBarros15}. Building up the sample size of such confirmed sources is crucial to aid our understanding of LyC photon escape from galaxies and of cosmic reionization.

In this paper we exploit the newly obtained UV imaging by the WFC3/UVIS camera on the \textit{Hubble Space Telescope} (\textit{HST}) from the \textit{Hubble} Deep UV (HDUV) Imaging Survey over the two GOODS/CANDELS-Deep fields (Oesch et al. 2016, submitted), along with data from the UVUDF survey \citep{Teplitz13,Rafelski15}. The HDUV filter set covers $\sim2500-3700~\angstrom$, and directly images LyC photons in $z\gtrsim2$ galaxies.
These UV data are combined with archival $HST$ imaging at longer wavelengths as well as spectroscopic redshifts from the literature to search for potential LyC candidates over the full redshift range $z=1.9$ to $z=4$, using a technique similar to the one presented in V15. This search also provides the basis for a future paper in which we will constrain the average escape fraction of star-forming galaxies at $z\sim2-3$.

This paper is structured as follows. In section \S2 we describe the imaging and the spectroscopic data used for the analysis. The methodology of our candidate search is outlined in \S3.  We present six candidate LyC emitters in \S4, calculate their $f_{esc}$ and situate them in the context of other efforts to understand LyC leakage in \S5, and finally summarize our findings while looking towards the future in \S6.

Throughout this paper, we adopt $\Omega_M=0.3, \Omega_\Lambda=0.7, H_0=70$ kms$^{-1}$Mpc$^{-1}$, i.e., $h=0.7$, largely consistent with the most recent measurements from Planck \citep{Planck2015}. Magnitudes are given in the AB system \citep{Oke83}.

\section{Data}
\label{sec:data}

\subsection{Photometry}
\label{sec:photometry}
As shown previously, any study of LyC emission at high-redshift requires data at excellent spatial resolution in order to avoid contaminating flux from foreground sources which lie close in projection along the line of sight \citep[e.g.][]{Vanzella10}.
Hence, in this paper we only analyze objects for which \textit{HST} images are available. The novel data that let us search for LyC candidates in the relatively unexplored redshift range of $z\sim2-3$ come from deep UV imaging (down to 27.5-28.0 mag at $5 \sigma$) of the GOODS-North and GOODS-South fields in the F275W and F336W bands acquired by the Hubble Deep UV (HDUV) Legacy Survey (GO-13871, see Oesch et al, 2016, submitted) including all the F275W data taken by the CANDELS survey \citep{candels}. Additionally, we include the previously released version 2 of the UVUDF images\footnote{\url{https://archive.stsci.edu/prepds/uvudf/}} \citep[][]{Teplitz13,Rafelski15}.

At $z\sim2-3$, the LyC is probed by the HDUV bands and we show in \S~\ref{sec:methods} how LyC leakage may be inferred from this photometric data.  Crucially, the HDUV survey's coverage area is a subset of that of the 3D$-$HST \citep{Skelton14, Brammer12, Momcheva16} and CANDELS \citep{candels, candels2} surveys, as well as the previous GOODS ACS imaging \citep[][]{Giavalisco04a}. This complementarity provides continuous multi-wavelength imaging (using ACS and WFC3) from the UV to the near-IR along with an abundance of grism redshifts (from 3D-HST), and facilitates the reliable calculation of UV-continuum slopes ($\beta$) and the derivation of physical properties of galaxies through spectral energy distribution (SED) fitting.

\subsection{3D-HST Grism Spectra and other Spec-Z}
\label{grism_spectra_data}
For a reliable LyC emitter search, we require accurate spectroscopic redshifts. Fortunately, the HDUV/GOODS fields have extended spectroscopic coverage from several surveys conducted over many years \citep[e.g.][]{Dawson01,Cowie04,Reddy06,Wuyts08,Yoshikawa10}. Most of the secure redshifts from these surveys are already compiled in the 3D-HST catalogs from \citet{Skelton14}, which we use for our analysis. We also harvest newly available spectroscopic redshifts from the VUDS \citep[DR1,][]{vuds} and MOSDEF \citep{mosdef} surveys. We ensure that only high-probability redshifts are included in our analysis (e.g., confidence class 3+ in the VUDS release).

Additionally, grism spectra from the 3D-HST survey are available for most of the sources that we analyze and they are particularly reliable when prominent emission lines are detected \citep{Momcheva16}. The WFC3/G141 grism used in the 3D-HST survey spans $1.1\mu m- 1.7\mu m$, and is perfectly situated to capture the distinctive [\ion{O}{3}]$\lambda\lambda4959,5007\angstrom$ doublet at $z\sim1.9-2.4$. Thus, the grism redshifts derived in this redshift window are anchored to well-detected emission lines and this gives us a sizable sample of sources for which the LyC is reliably located in the HDUV/F275W filter. Furthermore, the NIR spectra enable the analysis of emission line-ratios. In a narrow window around $z\sim~1.9-2.0$, both [\ion{O}{2}] and [\ion{O}{3}] fall in the G141 grism's spectral range and almost always, H$\beta$ is available along with [\ion{O}{3}], though blended with [\ion{O}{3}] given the very low spectral resolution of the grism.

\begin{figure*}[th]
\centering
\includegraphics[width = \textwidth]{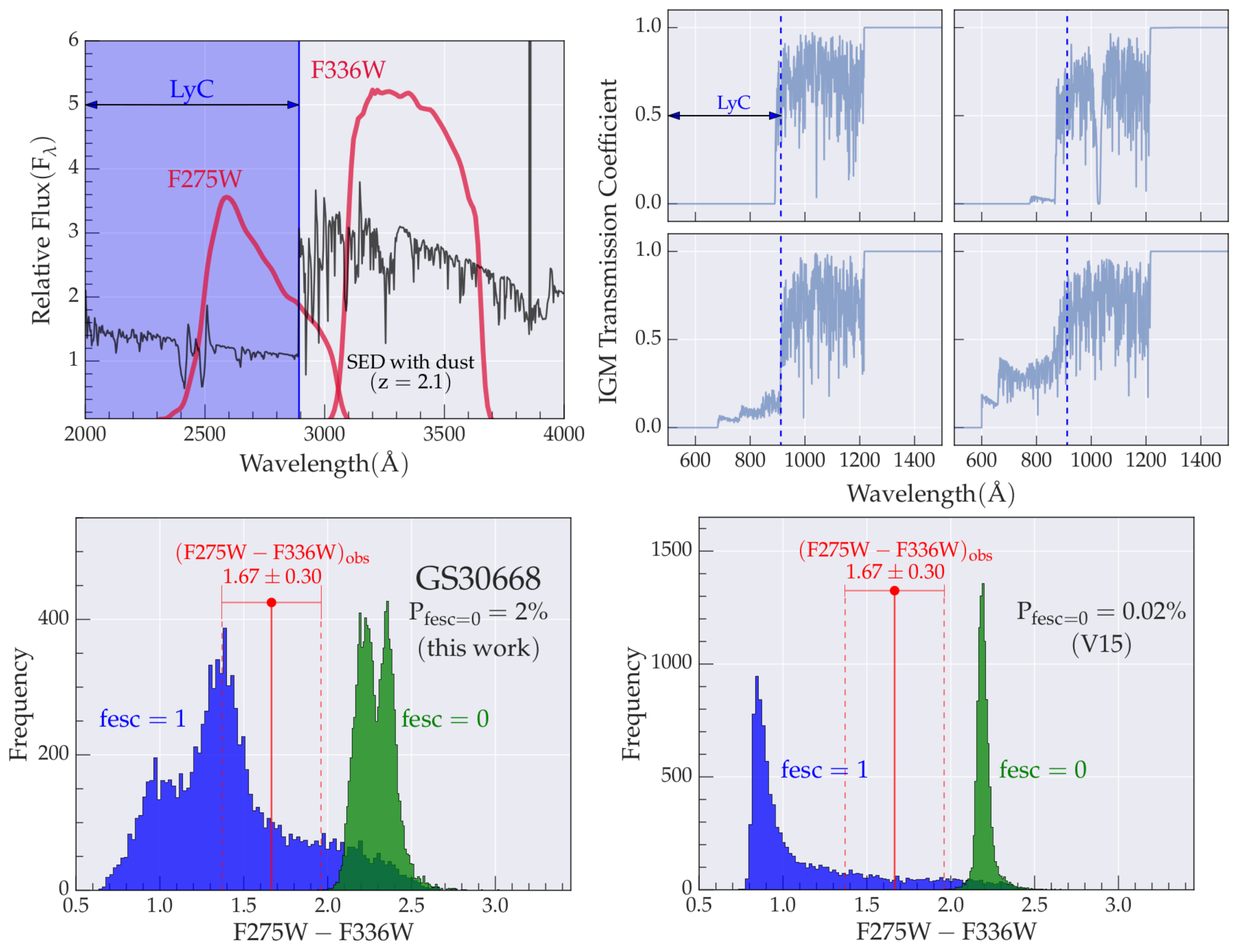}
\caption{Summary of the selection technique adopted in this paper illustrated with the first LyC candidate GS30668 ($z=2.172$). \textbf{Top left}: At GS30668's redshift, $>50\%$ of the F275W band is covered by the Lyman Continuum (shown as the shaded blue region) and hence LyC leakage may be inferred from the flux observed in that band. The SED shown is a BPASSv2 model SED with $\beta_{UV}$ and age consistent with GS30668, and E(B-V)=0.03 (using the dust extinction curve from \citet[][]{Reddy16a}). \textbf{Top right}: Four realizations of the IGM transmission curves toward a $z=2$ galaxy from \citet{Inoue14}'s Monte Carlo sampling with $\lambda = 912\angstrom$ indicated by blue dotted lines. 10,000 such curves are convolved with a Gaussian distribution of SEDs ($\mu =\beta_{Obs}, \sigma = \sigma_{\beta_{Obs}}$) in our method to compute the expected color distribution in the bottom left panel. \textbf{Bottom left}: The observed F275W-F336W color (red dot with errorbars) lies almost entirely blueward of the color distribution generated under the assumption that $f_{esc}=0$ (shown in green). The probability for $f_{esc}=0$ for GS30668 is $P(f_{esc}=0)=2\%$, which makes it an LyC candidate. Further, the actually measured color perfectly lines up with the distribution generated under the assumption that $f_{esc}=1$ (shown in blue), strongly suggesting a high value of $f_{esc}$ for this galaxy. \textbf{Bottom right}:  The same color distributions as on the left, but following the original V15 method, in which a single highly ionizing SED is used (selected from our BPASSv2 SED grid that uses the \citet[][]{Reddy16a} dust curve; see text) and there is no Monte Carlo treatment of the observed color. This method results in a tighter color distribution, but it also calculates a very low $P(f_{esc}=0)$ for GS30668. All six candidates found using our method of Gaussian SED distributions are also candidates according to the V15 method.}
\label{fig:method_summary}
\end{figure*}

\section{Methodology}
\label{sec:methods}

The HDUV filter set probes exclusively ionizing photons for galaxies at $z>2.4$ in F275W and at $z>3.1$ in F336W. For sources at somewhat lower redshift, the filters cover both ionizing and non-ionizing wavelengths (see Fig \ref{fig:method_summary} for an example of a $z=2$ galaxy). However, even at these lower redshifts, LyC emitters with non-negligible $f_{esc}$ can be identified by modeling the UV SED and estimating the contribution of non-ionizing photons to the filter flux of a given galaxy. In particular, V15 developed such a technique, which was successful in selecting \textit{Ion2} as a highly probable LyC candidate. \textit{Ion2} was subsequently followed up and confirmed with \textit{HST} imaging and until recently was the only spectroscopically confirmed LyC leaker at high redshift \citep[$z=3.212$, see][for more details on \textit{Ion2}]{deBarros15,Vanzella16}.

In this paper, we build on the V15 technique and adapt it to use a new dust curve and different SED models before applying it to the HDUV dataset. In brief, we fit the UV continuum slope of a galaxy to identify representative UV model SEDs which are then used to derive the expected color of that galaxy in the $HST$ filters straddling the LyC edge. The latter is done via a Monte-Carlo simulation of the IGM transmission representing 10,000 lines of sight. We then identify galaxies whose measured $HST$ colors are inconsistent with an escape fraction of $f_{esc}=0$, but indicate $f_{esc}>0$.

\begin{figure*}[tbp]
\centering
\includegraphics[width=0.72\textwidth]{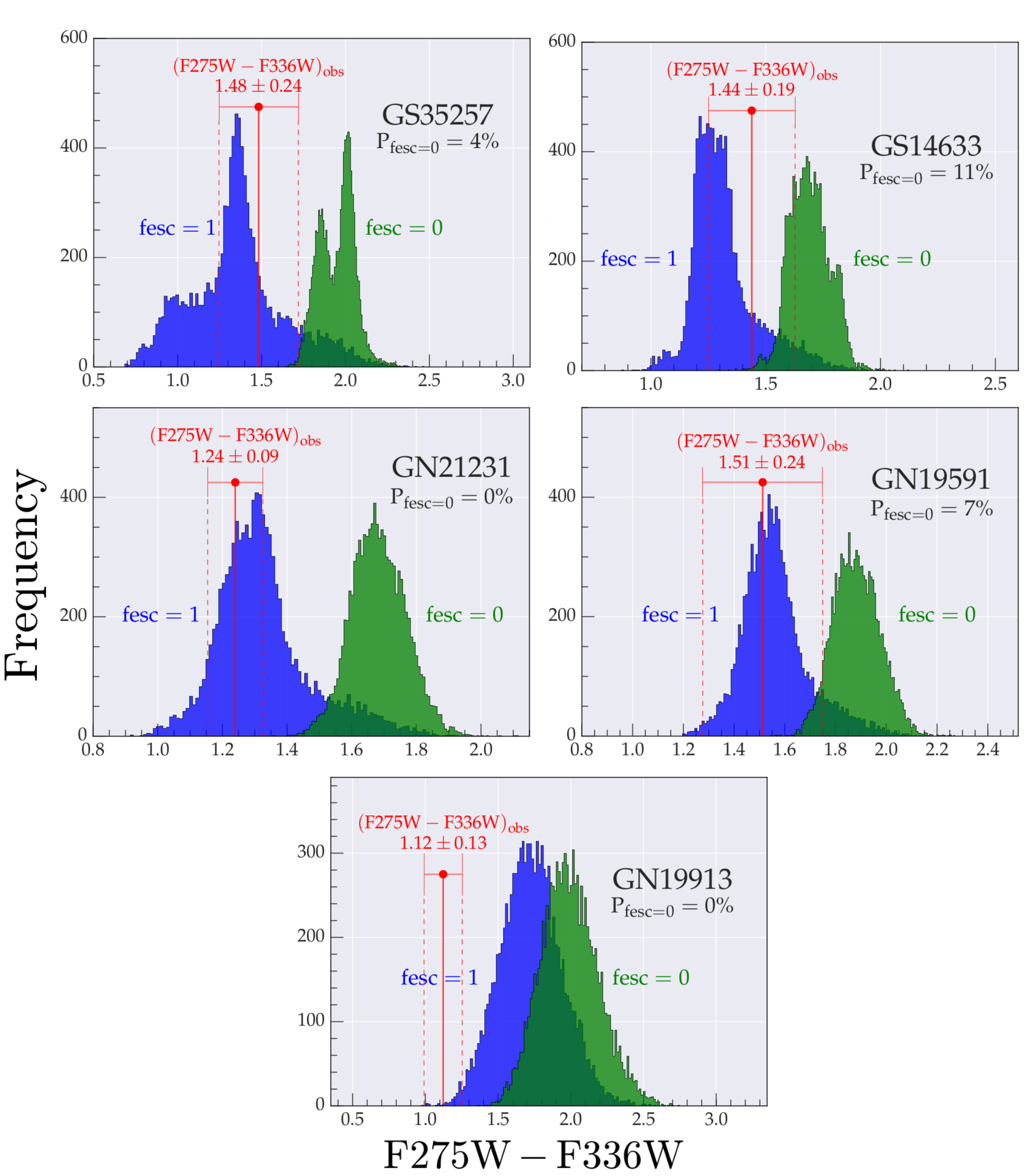}
\caption{Selection of LyC candidates from Monte-Carlo color distributions (see also Fig 1, bottom left panel). The galaxies whose simulated color distributions are shown here (all at $z\sim2$) are selected as LyC candidates since their observed F275W-F336W colors (indicated in red with errorbar) are inconsistent with the $f_{esc}=0$ distributions (shown in green), resulting in low $P(f_{esc}=0)$ values.}
\label{fig:combined_dists}
\end{figure*}

Explicit details of the selection procedure of sources with non-negligible $f_{esc}$ are described below:

\begin{enumerate}
  \item \textit{Input Galaxy Sample:} In order to obtain reliable results, we only apply our method to galaxies that have a secure spectroscopic redshift from the literature, as well as a reliable flux measurement (S/N$>3$) available in a filter that probes $>50\%$ of LyC photons (i.e., such that the central wavelength of the filter lies below rest-frame $912$\AA). The idea here is that if the galaxy is an LyC-leaker the measured filter-flux will contain a contribution from LyC photons that we can attempt to infer. A secure redshift is thus important, since it ensures that the LyC falls within a particular filter.

    \item \textit{SED grid:}    
    We assemble our SED grid using the latest Binary Population and Spectral Synthesis models (BPASSv2; Eldridge et al., in preparation) rather than the \citet[][henceforth BC03]{Bruzual03} models used in V15. This choice is motivated by the better performance of BPASS models in matching the spectral properties of star-forming galaxies at $z\sim2-3$ \citep[e.g.][]{Steidel16,Strom16}, as well as those of young, massive star clusters, which dominate the rest-frame UV region \citep{Wofford16}. The BPASS models are also more consistent with the intrinsic 900-to-1500$\angstrom$ flux density ratio inferred for $\sim$L* galaxies at $z\sim3$ \citep[][]{Reddy16b}.

    We use the fiducial BPASSv2 galaxy templates (IMF with a slope of -1.30 between 0.1 to 0.5$M_{\odot}$ and -2.35 from 0.5 to 100$M_{\odot}$) to which we self-consistently added nebular continuum and line emission to build an extremely fine template-grid that uniformly spans various ages, metallicities and magnitudes of dust extinction. The following parameter-space is covered in our grid: $Z/Z_{\odot} =0.05,  0.1 ,  0.15,  0.2 ,  0.3 ,  0.4 ,  0.5 ,  0.7 ,  1  , 1.5 ,  2 $; E(B-V) from $0-0.6$ in linear steps of $0.03$; age from 1Myr to 10Gyr in steps of 0.1 dex; and a constant star formation history. To account for dust attenuation we use the Calzetti attenuation curve \citep{Calzetti00} for wavelengths greater than $1500~\angstrom$. In the UV-region bluewards of $1500~\angstrom$ that is critical to our study, instead of simply extrapolating the Calzetti curve (as was done in previous studies, like V15), we use the newly derived dust curve from \citet{Reddy16a}. This new curve predicts a factor of $\sim2$ lower dust attenuation of LyC photons than the Calzetti curve for E(B-V)$\sim0.15$, typical for $\sim$L* galaxies. The difference in dust attenuation is smaller for bluer E(B-V).

     \item \textit{$\beta_{UV}$-IGM Monte Carlo color simulation and candidate selection:} In order to select SEDs that best represent a given galaxy, we fit the observed UV continuum slopes ($\beta_{obs}$) with the ones from the SEDs ($\beta_{SED}$).

    In the original V15 method, every galaxy is only matched to a single SED: from the grid described above, one selects the SEDs which satisfy $\beta_{Obs} - \sigma_{\beta_{Obs}}< \beta_{SED} < \beta_{Obs} + \sigma_{\beta_{Obs}}$ and $Age_{SED} <= Age_{Universe}$ at  $z_{spec}$ and from these, one picks the SED with the maximum $LLyC/L1500$ ($LLyC$ is calculated over $850-900\angstrom$). This results in the bluest simulated color (calculated for the LyC-containing filter and an adjacent filter) and is the most conservative to identify galaxies which are inconsistent with a zero escape fraction.

   In the method used in this work, instead of relying on a single extremely ionizing template, we account for the variance in potential ionizing fluxes of a given galaxy by sampling (10,000 times) from the SED library to produce a Gaussian UV continuum slope distribution centered at $\beta_{obs}$ and with $\sigma = \sigma_{\beta_{obs}}$. For each SED, we apply a realization of IGM attenuation drawn from 10,000 possible sightlines at the $z_{spec}$ of the given galaxy \citep[IGM transmission functions from][]{Inoue14}. The IGM-applied SEDs are used for a Monte-Carlo sampling of the expected color distribution of the galaxy in our HST filters by assuming $f_{esc}=0$. In the V15 method, the color distribution arises from the convolution of the single extremely ionizing template with the 10,000 IGM sightlines and filters.

If the color distribution, which has no LyC photons contributing to it (i.e., assuming $f_{esc}=0$), falls largely redward of the observed color, the galaxy is an LyC candidate since even in transparent IGM sighlines LyC photons would be required to reproduce the observed color.
This idea is formalized in terms of a probability (Equation 1 from V15):
  $$P(f_{esc}=0) = N_{color}/N_{total} $$
  where $N_{total}=10^9$ and $N_{color}$ is the number of IGM-$\beta$-$Color_{obs}$ realizations for which $Color_{simulated} > Color_{obs}$. 
  We treat $Color_{obs}$ as a Gaussian distribution with width corresponding to the photometric scatter, and sample 100,000 times from this distribution for each of the 10,000 IGM lines of sight, giving $N_{total}=10^9$. LyC candidates are the galaxies for which this probability $P(f_{esc}=0)$ is low. We repeat the calculation of color histograms under the assumption of $f_{esc}=1$ for a consistency check (see blue histograms in Figs. \ref{fig:method_summary}, \ref{fig:combined_dists}).

\end{enumerate}

 While both the V15 method and our method can be used to select LyC leakers, our method can also be used to constrain $f_{esc}$ since it accounts for the entire diversity in the UV SED of the galaxy as well as the variation in the IGM transmission (see \S5.1). In general, the two methods identify the same candidate LyC emitters. However, there are some marked differences. For instance, in Figure \ref{fig:method_summary}, we show the expected color distributions computed with both methods for the first candidate selected in the HDUV dataset (GS30668). This source is a high probability candidate with $P(f_{esc}=0)=2\%$ (our method) and $P(f_{esc}=0)=0.02\%$ (V15). However, the color distributions computed using our method are much wider, since we do not adopt just a single SED but a distribution of SEDs according to the variance in the $\beta_{obs}$ of the galaxy. In the case of V15, the same SED is used for all realizations and the color distribution is only a reflection of the variance of IGM transmission functions.

\begin{figure*}[th]
\centering
\includegraphics[width=0.77\textwidth]{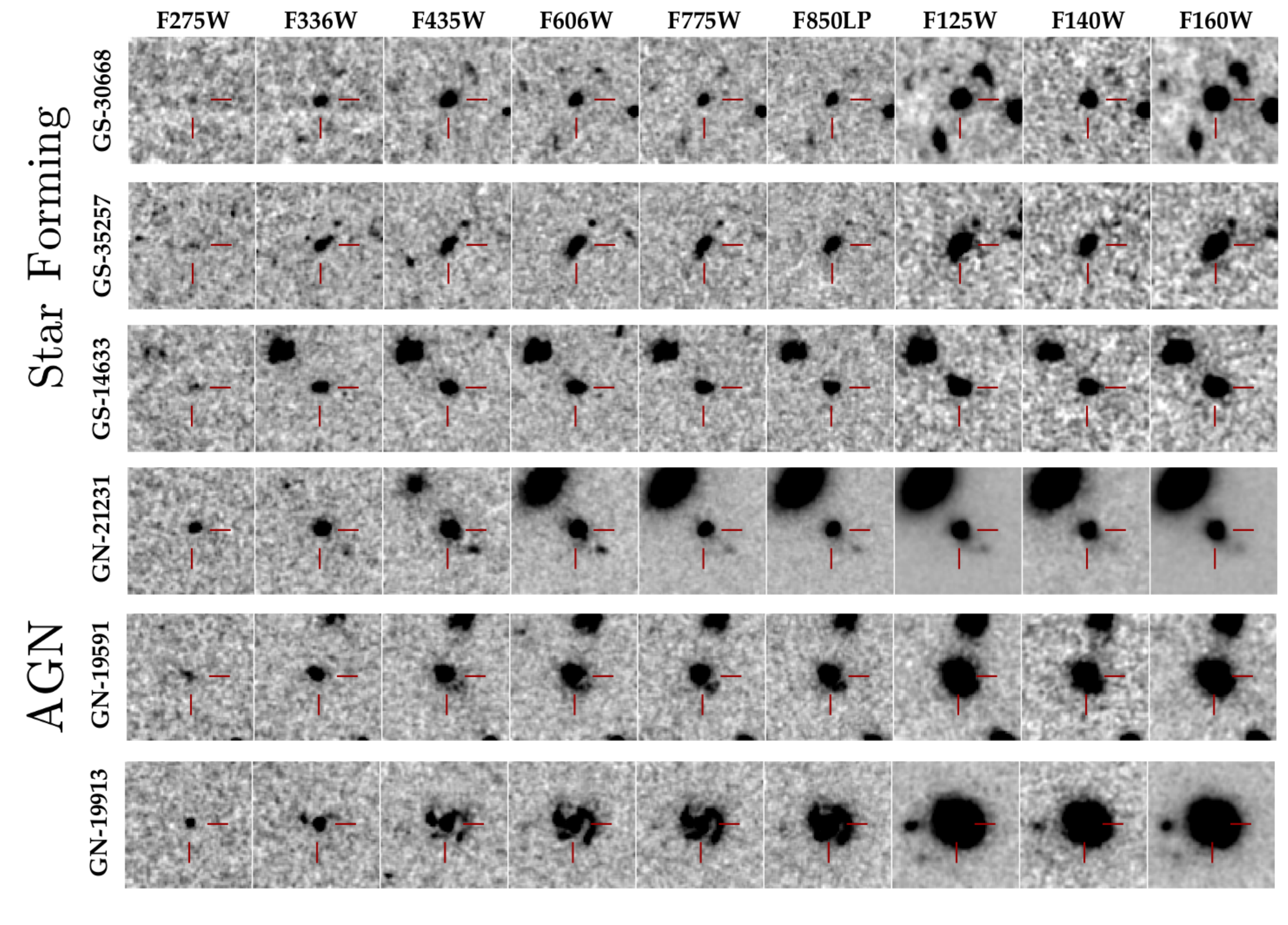}
\caption{3\arcsec HST postage stamp images of LyC candidates. The F275W and F336W images (first two columns) are the new data acquired by the HDUV survey, while the rest are from HST-GOODS archival imaging (see \S\ref{sec:photometry}). These high resolution, multiwavelength images allow us to conclusively rule out flux contamination from neighbouring sources. For instance, in the case of GN19591  we investigate a potential interloping clump towards the bottom right of the central source (visible in F435W, F606W, F775W), but we conclude that this does not pose a problem (see \S\ref{sec:candidate_agn})--such a check wouldn't be possible at ground-based resolution. In terms of morphology, our candidates are in general compact. Two curiosities are the AGN GN21231 and GN19591, whose F275W detections (which includes hypothesized LyC leakage) appear to be extended and not concentrated on a central point source (see the AGN GN19913 for comparison which is barely resolved in F275W) which may implicate stars instead of the active nucleus as the origin of LyC flux (see \S\ref{sec:candidate_agn} for details).}
\label{fig:stamps}
\end{figure*}

\section{Candidate Lyman Continuum Leakers}

The procedure outlined in \S3 allows us to identify likely LyC emitter sources in the redshift range $z=1.9-4$. The upper limit in the redshift range is primarily imposed by the decreasing IGM transmission to higher redshift, which makes it virtually impossible to directly probe LyC photons at $z>4$ based on broad-band images. Below the lower redshift limit, our bluest filter (F275W) contains $<50\%$ of the LyC. We thus applied our procedure to all sources listed in the 3D-HST GOODS catalogs with a secure redshift in the range $z=1.9-4$ for which the relevant photometry was available.

For a source to qualify as a LyC emitter candidate we set the following criteria: \textit{(I)} $3 \sigma$ detection in the LyC containing band, $5 \sigma$ detection in the adjacent redward band used to calculate the color distribution, and S/N greater than $5$ for the calculated color, \textit{(II)} a robust $\beta$ slope fit using flux measurements from at least three bands, \textit{(III)} clean morphology in all available $HST$ images to rule out contamination from low-$z$ interlopers with chance projections, and \textit{(IV)} $P(f_{esc}=0) < 15\%$.

Using these criteria, we identify six candidates in the HDUV+UVUDF survey area. Their color histograms are shown in Figure \ref{fig:method_summary} for the first source and Figure \ref{fig:combined_dists} for the remaining five, while their basic properties are listed in Table 1. In principle, we extended our search up to $z\sim4$, however, all the six candidates lie at $z\sim2$ and they were selected based on the F275W-F336W color. This is not necessarily surprising, given that most of our input spectroscopic redshifts from the literature, and in particular, from the 3D-HST grism data lie at $z\sim2$. In total, our input galaxy sample with reliable spectroscopic redshifts contained 1124 galaxies.

\begin{figure*}[th]
\centering
\includegraphics[width=0.7\textwidth]{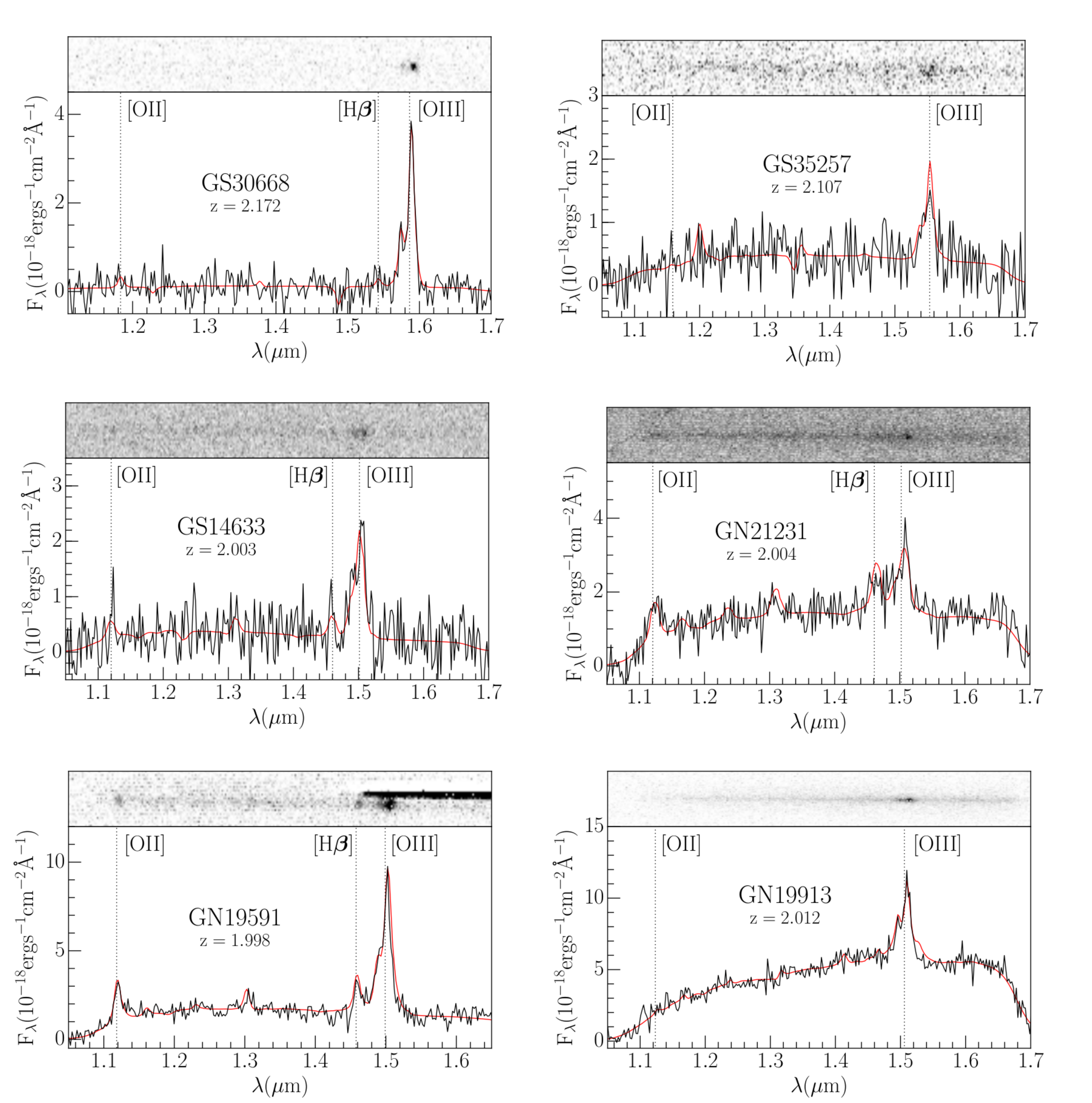}
\caption{3D-HST grism spectra of LyC candidates. In each panel the top strip contains the 2D-spectrum of the source and the bottom part shows the extracted 1D-spectrum (black) along with a best-fit model (red) (see \citet{Momcheva16} for details about the 3D-\textit{HST} pipeline). In all the spectra, the distinctive merged [\ion{O}{3}] doublet (\ion{O}{3}4959 $\angstrom$+ \ion{O}{3}5007$\angstrom$) is unambiguously detected. On the other hand, the OII and $\mathrm{H{\beta}}$ detections are in general tentative. Deeper, high resolution spectra are required to better constrain [\ion{O}{3}]/[\ion{O}{2}] and $\mathrm{H{\beta}}$/$\beta_{UV}$, two promising indirect methods to infer the escape fraction of galaxies. }
\label{fig:grism_stack}
\end{figure*}

We pay careful attention to redshift quality, since our entire selection procedure hinges on secure redshifts. All our candidates have robust 3D-HST grism spectra with well-detected emission lines (shown in Figure \ref{fig:grism_stack}), and we use the associated $z_{grism}$ measurements in our analysis. Additionally, spectroscopic redshifts were already available from the literature for four of the sources which corroborate the $z_{grism}$ (\citealt[][]{Reddy06} for GN19591, GN19913 and GN21231; \citealt[][]{Kirkpatrick12} for GN21231; \citealt[][]{Trump11} for GS30668).

We investigated the sources in detail and tested whether they show any sign of AGN activity which could contribute to the ionizing photons, or whether they had any nearby neighbors that could contaminate the UV color measurements.
We split the sample in two classes: (1) star-forming galaxy candidates, and (2) likely active galactic nuclei (AGN). These classes are discussed separately in the following sections.

\subsection{Star-Forming Galaxy Candidates}
\label{sec:sf_candidates}
The first three of our candidates are the most convincing as they show no signs of AGN activity and no nearby, potentially contaminating galaxies. These are GS30668, GS35257, and GS14633. These sources are not detected in the deepest available Chandra GOODS images and associated CANDELS-matched catalogs that include even extremely faint X-ray sources \citep{Cappelluti16}. 
For these three source we can further examine the Mass-Excitation diagram \citep[][]{Juneau14}, since reliable [\ion{O}{3}] and H$\beta$ measurements are available. GS30668 and GS14633 are within the star-forming region of the diagram while GS35257 lies on the border separating AGN and star-forming galaxies. The compact and isolated objects in this section are likely galaxies where ionizing radiation escapes similar to the source $Ion2$ identified in V15 using a technique similar to the one we have adopted here.

\subsection{AGN}
\label{sec:candidate_agn}
Two sources in our sample are known AGN$-$GN21231 and GN19913, which have been well-studied in the literature (GN19913: e.g. \citealt[][]{Smail04, Bluck11}; GN21231: e.g. \citealt[][]{Evans10,Kirkpatrick12}). In addition to them, we classify GN19591 as a likely AGN based on a strong $Spitzer$/MIPS and even a Herschel detection, and the hint of a power law SED across the $Spitzer$/IRAC bands. However, we note that GN19591 is not detected in the Chandra Deep Field North (2 Msec) images, and has spectral features that could easily belong to a star-forming galaxy (e.g. narrow width of the \ion{O}{3} line,  [\ion{O}{3}]/[\ion{O}{2}]$\sim$2.2, [\ion{O}{3}]$\lambda5007\angstrom/\mathrm{H}_{\beta} \sim 3.2$).

GN21231 was identified as an AGN by \citet[][]{Kirkpatrick12} and was found to display the distinctive 9.7 $\mu m$ silicate line in its IR spectrum, which indicates a large column of dust along the line of sight. This massive amount of dust is perhaps why GN21231 is heavily obscured and not detected in Chandra images. The dust probed by the 9.7 $\mu m$ line is concentrated in the innermost, central 2pc of the galaxy \citep[][]{Kohler10}, however. It is thus fair to conjecture that the ionizing flux from the active nucleus in GN21231 is largely suppressed by the dust torus around it. Thus, any LyC flux we detect probably originates from the star forming parts of the galaxy. This hypothesis is supported by our FAST SED fit which yields E(B-V)$=0$, indicating that while the central nucleus may be dusty, the rest of the galaxy is not. The morphology of the ionizing radiation in the F275W filter, which is extended and not concentrated on a central point-source, is another indication that the LyC flux originates from stars.

Similar to GN21231, GN19591 which is not detected in Chandra 2 Msec images might host a heavily obscured AGN. Based on the extended, resolved LyC radiation in the F275W image, it is likely that star-forming regions are responsible for the ionizing flux, however. The similarity of morphologies in the F336W and F275W images is further evidence of the star-forming regions being correlated with LyC emission. Viewed together, these two obscured AGN candidates suggest that AGN may play an important role in clearing the ISM and facilitating the escape of LyC photons emitted by stars in these galaxies.

In contrast, the situation in the AGN GN19913 is clearer, and indicates
that it is not useful for a study of LyC escape from stars. The UV
morphology of GN19913 is concentrated on a central point source consistent
with the ionizing photons originating from the AGN.

\begin{figure*}[th]
\centering
\includegraphics[width=0.92\textwidth]{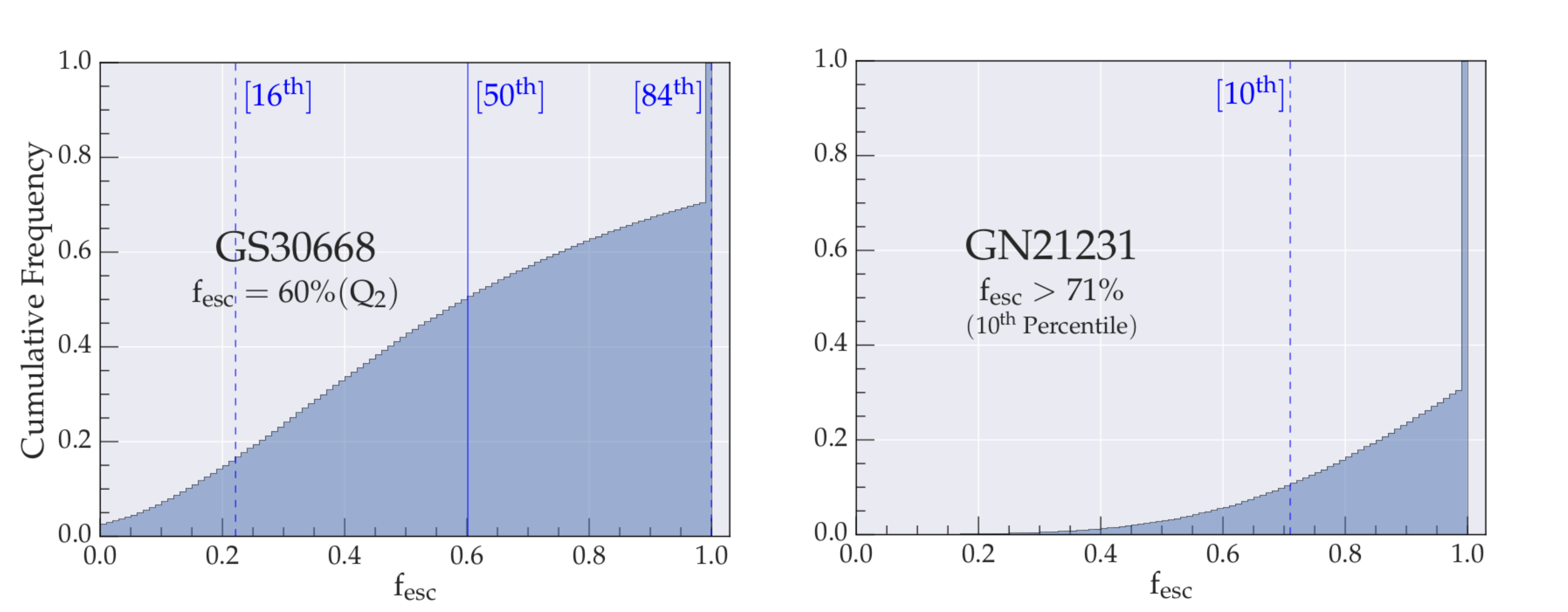}
\caption{Example $f_{esc}$ distributions for two of our candidates. The cumulative $f_{esc}$ distributions based on 100,000 $f_{esc}$ realizations (ten observed color samplings per IGM line of sight) are shown, with percentiles for the distribution indicated in blue text. In the left panel, for star-forming galaxy GS30668, we state the median $f_{esc}$. On the right, more than half of the $f_{esc}$ values calculated for the source GN21231 are truncated to 1. In this case, we state a lower limit equal to the $\mathrm{10^{th}}$ percentile. See Table \ref{summary_table} for similarly calculated $f_{esc}$ values for all our candidates.} 
\label{fig:fesc_dists}
\end{figure*}

\begin{table*}[th]
\centering
\caption{Summary of Lyc Candidates}
\label{summary_table}
\begin{tabular}{lcccccc}
\hline \hline
\multicolumn{1}{l}{} & \multicolumn{3}{c}{Star-Forming Galaxies} & \multicolumn{3}{c}{Active Galactic Nuclei}\\
\noalign{\smallskip}
 & GS30668 & GS35257  & GS14633 & GN21231  & GN19591 & GN19913\\
\noalign{\smallskip}
\hline
\noalign{\smallskip}
$\mathrm{R.A.}$  &  $3:32:35.47$  &  $3:32:24.93$  &  $3:32:46.46$\ &  $12:36:46.74$  &  $12:36:48.31$  &  $12:36:35.6$  \\[4pt] 
$\mathrm{Dec.}$   &  $-27:46:16.89$  &  $-27:44:51.61$  &  $-27:50:36.64$ \ &  $+62:14:45.97$  &  $+62:14:16.64$  &  $+62:14:24.0$  \\[4pt] 
$\mathrm{Redshift}$\tablenotemark{$a$}   &  $2.172_{-0.003}^{0.001}$  &  $2.107_{-0.003}^{+0.002}$  &  $2.003_{-0.002}^{+0.001}$\ &  $2.004_{-0.002}^{+0.002}$  &  $1.998_{-0.003}^{+0.001}$  &  $2.012_{-0.002}^{+0.001}$  \\[4pt] 
$\mathrm{f_{esc}} [\%]$ \tablenotemark{$b$}  &  $60^{+40}_{-38}$  &  $72^{+28}_{-48}$  &  $62^{+38}_{-51}$\ &  $>71$  &  $>13$  &  $\sim100$  \\[4pt] 
$\mathrm{UV~ slope}~ \beta ~~(f_\lambda\propto \lambda^\beta)$  &   $-2.23\pm0.08$  &   $-1.93\pm0.07$   &  $-1.92\pm0.01$ \ &   $-1.86\pm0.07$  &   $-1.27\pm0.07$  &  $-1.11\pm0.41$  \\[4pt] 
$\log \mathrm{M_{gal}/M_\odot}$\tablenotemark{$c$}  &  $9.07_{-0.06}^{+0.05}$ &  $9.37_{-0.03}^{+0.00}$ &  $9.23_{-0.01}^{+0.00}$\ &  $10.44_{-0.03}^{+0.00}$ &  $9.99_{-0.03}^{+0.12}$ &  $11.3_{-0.06}^{+0.00}$ \\[4pt] 
$\log \mathrm{SFR} [\mathrm{~M_\odot~\mathrm{yr}^{-1}}]$\tablenotemark{$c$}  &  $0.24_{-0.04}^{+0.02}$  &  $0.38_{-1.33}^{+0.0}$  &  $-0.56_{-1.39}^{+0.00}$\ &  $1.06_{-0.00}^{+0.01}$  &  $0.92_{-0.00}^{+1.29}$  &  $1.55_{-0.41}^{+0.00}$  \\[4pt] 
$\log \mathrm{sSFR} ~[\mathrm{yr}^{-1}]$\tablenotemark{$c$}   &  $-8.83_{-0.04}^{+0.07}$  &  $-8.99_{-1.31}^{+0.00}$  &  $-9.79_{-1.38}^{+0.00}$\ &  $-9.38_{-0.00}^{+0.04}$  &  $-9.07_{-0.00}^{+1.30}$  &  $-9.75_{-0.35}^{+0.00}$  \\[4pt] 
$\log \mathrm{Age[yr]}$\tablenotemark{$c$}   &  $8.9_{-0.2}^{+0.1}$  &  $8.0_{-0.1}^{+0.1}$  &  $8.0_{-0.0}^{+0.0}$\ &  $9.5_{-0.0}^{+0.0}$  &  $7.7_{-0.1}^{+0.6}$  &  $9.5_{-0.0}^{+0.0}$  \\[4pt] 
$\mathrm{E(B-V)}$\tablenotemark{$c$}   &  $0.0_{-0.0}^{+0.0}$  &  $0.07_{-0.03}^{+0.01}$  &  $0.02_{-0.02}^{+0.00}$\ &  $0.0_{-0.00}^{+0.01}$  &  $0.22_{-0.00}^{+0.10}$  &  $0.17_{-0.10}^{+0.01}$  \\[4pt] 
[\ion{O}{3}]/[\ion{O}{2}]\tablenotemark{$d$}  &   $9.47\pm3.81$  &   $>5.5$   &  $3.23\pm1.41$\ &   $0.81 \pm 0.12$  &  $2.23\pm0.17$  &  $10.33 \pm 5.43$  \\[4pt] 
$\mathrm{EW_{rest}(}$[\ion{O}{3}]$\mathrm{)[\angstrom]}$\tablenotemark{$d$}  &   $1211\pm55$\tablenotemark{$e$}  &   $168 \pm 31$   &  $501\pm89$\ &   $102 \pm 10$  &  $384\pm22$  &  $94 \pm 4$  \\[4pt] 
$\mathrm{EW_{rest}(H\beta)[\angstrom]}$  &   $130\pm45$\tablenotemark{$e$}  &   $<19$   &  $95\pm38$\ &   $49 \pm 9$  &  $97\pm10$  &  $13 \pm 3$\\[4pt]
\hline
\noalign{\smallskip}
\end{tabular}
\begin{tablenotes}
\item[](a) Redshift is the 3D-\textit{HST} $z_{grism}$ for all sources.
\item[](b) Median of $\mathrm{f_{esc}}$ distribution with $\mathrm{16^{th}}$ and $\mathrm{84^{th}}$ percentile error-bars. When $>50\%$ of the distribution is truncated, we state the $\mathrm{10^{th}}$ percentile as a lower limit  (see \S~\ref{sec:measuring_fesc}).
\item[](c) Derived using FAST \citep[][]{Kriek09} and BC03.
\item[](d) [\ion{O}{3}] refers to the merged doublet i.e. [\ion{O}{3}]4959 $\angstrom$+ [\ion{O}{3}]5007$\angstrom$.
\item[](e) The 3D-\textit{HST} pipeline overestimates equivalent widths when the continuum detected by the grism is very faint, like in the case of GS30668 (see Figure \ref{fig:grism_stack}). In such a case we use the grism line flux along with a continuum extrapolation from EAZY \citep[][]{Brammer08} to calculate the equivalent width.
\end{tablenotes}
\end{table*}

\section{Discussion}
\label{sec:discussion}

\subsection{Probability Distributions of the LyC Escape Fraction}
\label{sec:measuring_fesc}

We can use the color histograms presented in Figures \ref{fig:method_summary} and \ref{fig:combined_dists} to estimate probability distributions of the absolute $f_{esc}$ for each source. In particular, for each realization of the IGM transmission we can compute the required escape fraction to bring the simulated $F275W-F336W$ color into agreement with the observed color (which is treated as a Gaussian distribution, as described in \S\ref{sec:methods}).

For a particular sight line, if the observed color is bluer (redder) than the simulated $f_{esc}=1$ ($f_{esc}=0$) color we truncate $f_{esc}$ to 1 (0). For every candidate we have a billion estimates of $f_{esc}$ (10$^5$ samplings from the observed color Gaussian for each of the 10,000 IGM realizations), based on which we compute the cumulative distribution function of $f_{esc}$ (see Fig \ref{fig:fesc_dists}).
In Table \ref{summary_table} we summarize these measurements. We state the median $f_{esc}$ with the $16^{\textrm{th}}$ and $84^{\textrm{th}}$ percentiles as error bars. When $>50\%$ of the $f_{esc}$ estimates for a source are truncated at $f_{esc}=1$, we state the $10^{\textrm{th}}$ percentile as a lower bound. 

The estimated $f_{esc}$ values for our candidates range from $\sim~60\%$ to $\sim~100\%$\footnote{Note that $f_{esc}=100\%$ or higher is either unphysical or can be excluded due to our detection of strong emission lines in these objects (Fig. \ref{fig:grism_stack}). Such high inferred values rather reflect our limited knowledge of the intrinsic UV SEDs below the Lyman limit.}. Such high escape fractions are expected, given our selection procedure in which we set the stringent criterion of $P(f_{esc}=0) < 15\%$. This is also in agreement with the findings of $Ion2$, which was selected as an LyC candidate by the V15 method, and was later confirmed to have an $f_{esc}\gtrsim50\%$ through follow-up imaging with $HST$ \citep{Vanzella16}. 
Note that it is likely that many sources in our parent input sample show significant escape fractions, but they are missed in our selection since we can not reliably separate the two color histograms. A future paper in preparation will address the average escape fraction of galaxies in the HDUV fields. In general, our findings are in broad agreement with previous studies that found only a small fraction of galaxies to show high $f_{esc}$. This could be explained by generally high gas covering fractions with few clear sightlines out of galaxies or similarly that the escape of ionizing photons is a stochastic process, with short periods of time of high $f_{esc}$ \citep[e.g.][]{Wise14,Cen15}.

The two confirmed AGN in our sample show particularly high $f_{esc}$ (GN21231 $>71\%$ [$10^{\mathrm{th}}$ percentile]; GN19913 $\sim100\%$), with the caveat that we have not included AGN templates in deriving these values. For comparison,  \citet[][]{Matthee16} report two AGN at $z\sim2$ with $f_{esc}$ $<60\%$ and $<75\%$. For their four LyC leaking AGN at $z\sim3$, \citet[][]{Micheva16} derive $f_{esc}$ values of $<0.016, 0.73\pm0.30, <0.007,$ and $0.31\pm0.27$ by allowing for median IGM attenuation. At $z\sim 3.8$, \citet[][]{Cristiani16} use a large sample of QSOs to derive a mean escape fraction of $75\%$.

\subsection{Comparison with confirmed high-$z$ LyC leakers}
\label{comparison_highz}

Here we provide a short comparison of our LyC emitter candidates with the three previous, confirmed LyC sources at $z>2$: $Ion2$ ($z=3.21$), Q1549-C25 ($z=3.15$) and  MD5b ($z=3.14$). 
In particular, we estimated several physical parameters for our LyC leaker candidates based on the broad-band photometry and SED fitting using FAST. These are summarized in Table \ref{summary_table}. In general, our galaxies have $\sim L*_{UV}$ at $z\sim2$ \citep[]{Reddy09}. 

$Ion2$ and GS30668 share many remarkable similarities. Both display an extreme EW([\ion{O}{3}]$_{4959+5007}$ + H$\beta$)  ($Ion2$: $\sim 1600 \angstrom$; GS30668:$\sim 1350 \angstrom$) and [\ion{O}{3}]/[\ion{O}{2}] $\gtrsim 10$. Even the multi-band fitted $\beta_{UV}$ ($Ion2:-2.2\pm0.2$; GS30668:$-2.2\pm0.1$), EW($H\beta$) ($Ion2$:112$\pm$60$\angstrom$; GS30668:130$\pm$45$\angstrom$) and E(B-V)$=0$ for these sources resemble each other.

Broadly speaking, our candidates display little to no dust extinction, consistent with $Ion2$, Q1549-C25 and MD5b. The exception is GN19591 with E(B-V)=$0.22_{-0.00}^{+0.10}$. This trend supports the idea that dust attenuation is not conducive to the escape of LyC radiation.
Furthermore, similar to previous LyC leakers (in particular MD5b), the star-forming galaxies in our sample are generally very young ($\sim$ 50-160 Myrs). GS30668 is the exception, with an age of $\sim 800$ Myr (once again, consistent with $Ion2$'s reported age).

Our sources thus provide some evidence that LyC emission generally occurs in young galaxies with little dust extinction and thus blue SEDs, or in sources whose ISM is highly excited and who show very strong rest-frame optical emission lines (such as $Ion2$) \citep[consistent with][]{Jones13, Wofford13, Borthakur14, Alexandroff15, Rivera-Thorsen15, Trainor15, Dijkstra16, Reddy16b, Nakajima16}. However, a more systematic study of the average escape fractions of galaxies in the HDUV field will have to be performed to better connect the physical properties that lead to LyC emission with significant $f_{esc}$ from galaxies.

\subsection{The Peak Escape Fraction as a Function of Redshift}

\begin{figure}[tbp]
\centering
\includegraphics[width=1.05\linewidth]{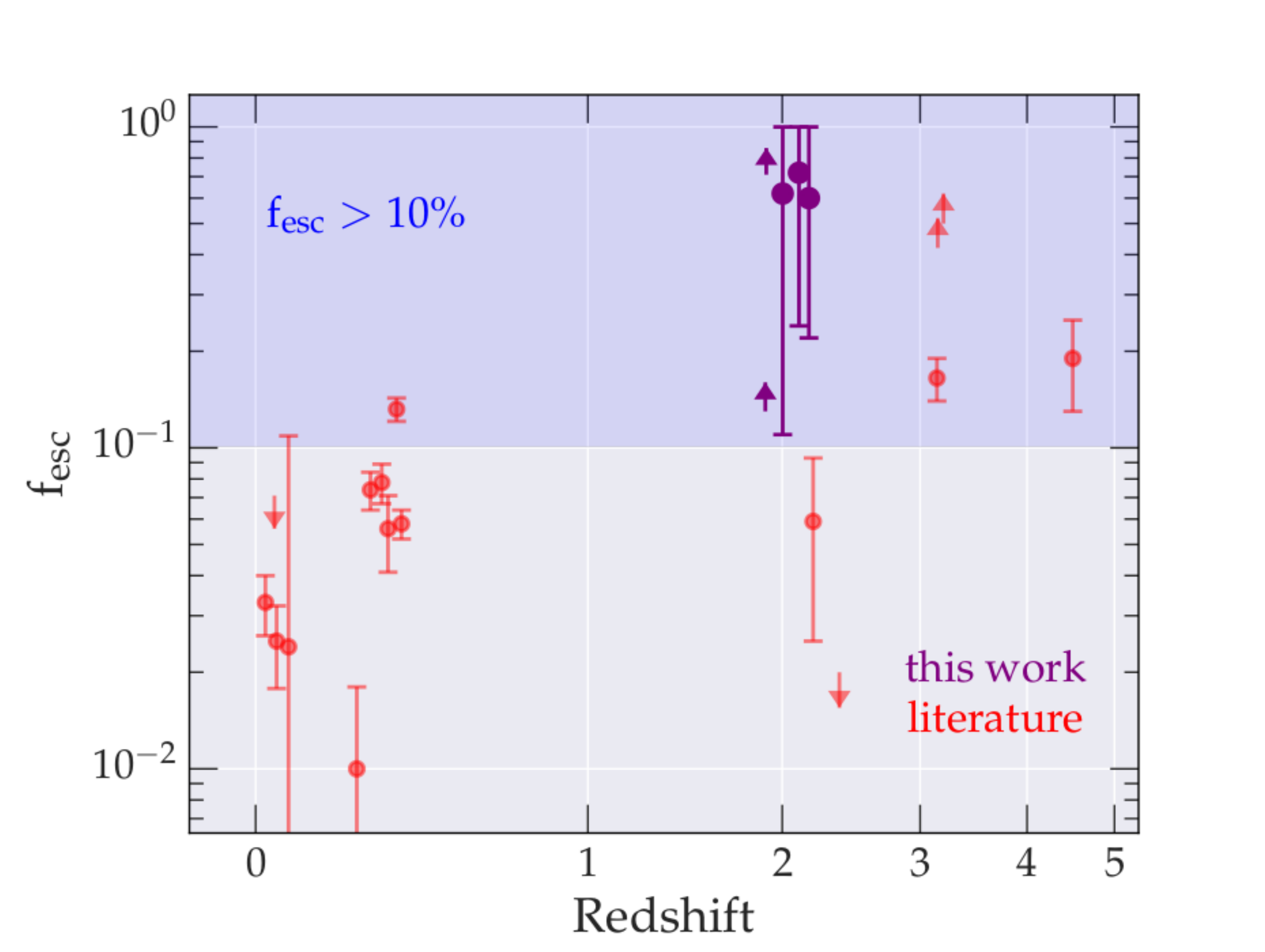}
\caption{A compilation of absolute $f_{esc}$ measurements for star-forming sources reported in the literature. The red points include direct detections from galaxies at low redshift \citep[][]{Leitet11,Borthakur14,Izotov16a,Izotov16b,Leitherer16} as well as detections or limits from $z>2$ sources using different methods \citep[][]{Mostardi15,Shapley16,Vanzella16,Vasei16,Matthee16,Leethochawalit16}. The candidate sources studied in this paper (shown in purple) occupy the relatively unexplored $z\sim2$ region in redshift space, and they double the number of direct high-$z$ $f_{esc}$ detections. The redshift of some sources was slightly offset for clarity. The shaded area in the upper half of the graph represents $f_{esc}>10\%$, a necessary condition for star-forming galaxies to drive reionization. While the population-average escape fraction at $z>2$ still has to be measured reliably, it is clear that at least some of the few individually detected sources at high redshift satisfy this criterion. Only one such source is currently known at $z<0.5$, hinting at a possible evolution of the maximally achievable escape fraction as a function of cosmic time. Note, however, that the $z>2$ sources were selected based on their high $f_{esc}$, and that the absolute value of the $f_{esc}$ measurements depend on the exact assumptions made (see text).
}
\label{fig:summary}
\end{figure}

It is interesting to put the star-forming galaxies identified in this paper in a broader context. Figure \ref{fig:summary} shows a compilation of absolute $f_{esc}$ measurements for star-forming sources reported in the literature. These include recent direct detections for individual galaxies at low redshift \citep[][]{Leitet11,Borthakur14,Izotov16a,Izotov16b,Leitherer16} as well as detections or limits from individual $z>2$ sources \citep[][]{Mostardi15,Shapley16,Vanzella16,Vasei16} or averages from subsamples of galaxies at $z>2$ \citep[][]{Matthee16,Leethochawalit16}. The plot does not show population average escape fractions, which still have to be measured reliably based on large samples of galaxies with $HST$ imaging in the future \citep[e.g.][]{Siana15}. In addition to our star-forming candidates, we also show the two AGN for which we have good evidence that the ionizing photons we detect are emitted by star-forming regions (GN21231 and GN19591).

While the lower redshift sources that are now being detected as LyC emitters typically still show a relatively low escape fraction of $<15\%$, a significant fraction of the high-redshift detections reach $f_{esc}\gtrsim50\%$. 
Given that most of the high-redshift points included in Figure \ref{fig:summary} were selected based on their high $f_{esc}$, it is clear that they are not likely representative of the average galaxy at these redshifts. However, the fact that several galaxies with likely $f_{esc}>50\%$ at $z>2$ have been found, while no such sources have (so far) been seen at $z<2$, hints at a possible evolution of the maximally achievable escape fraction from galaxies as a function of cosmic time \citep[see also][]{Inoue06}.

Note that the various derivations of $f_{esc}$ in the literature use different assumptions (e.g., SED frameworks, IMF, median/mean stacking etc.) which will affect the absolute values that are reported and shown in Fig \ref{fig:summary}. For instance, SED models which include binary stellar populations (like BPASSv2, used in this work) produce a larger number of ionizing photons and thus lead to lower $f_{esc}$ values compared with models like BC03 that have often been used in the past literature. The magnitude of this effect depends on the exact assumptions, but it is of order $\sim2-3\times$. This is still smaller than the order of magnitude difference seen between the reported $f_{esc}$ values found for low- and high-redshift sources.

Another caveat for the above conclusion of an evolving peak escape fraction is that we do not have a complete sampling of lower redshift LyC sources. While LyC photons can be directly observed at $z\gtrsim1$ through UV imaging surveys, the current lower redshift LyC emitters are all obtained through targeted, individual follow-up observations with UV spectrographs. Even though the most likely LyC candidate sources are typically followed up, it is not guaranteed that no sources with $f_{esc}>20\%$ exist, and it will be important to continue to search for these with future observations.

\subsection{Linking $f_{esc}$ to $z>5$ Observables}
The opacity of the IGM prevents any direct measurement of $f_{esc}$ beyond $z\gtrsim4.5$. So in order to study $f_{esc}$ in the Epoch of Reionization, we need to link it to quantities that may be measured at very high redshifts.
Several such indirect indicators of $f_{esc}$ have been discussed in the literature, including: (1) the line ratio [OIII]/[OII] which potentially traces density-bounded HII regions \citep[e.g.,][]{Jaskot13,Nakajima14,Faisst16}, (2) the strengths of nebular emission lines such as H$\beta$ compared with the total star-formation rate \citep{Zackrisson13,Zackrisson16}, (3) the shape of the Ly$\alpha$ line profile \citep{Verhamme15,Verhamme16}, or (4) the absorption strength of low-ionization lines and Lyman series lines which are related to the covering fraction of absorbing gas \citep[e.g.,][]{Heckman11, Leethochawalit16, Reddy16b}.
With the limited data we already have on our candidates, we can discuss the first two indicators, which we do in the following sections.

\subsubsection{[\ion{O}{3}]/[\ion{O}{2}]}
It has been shown that $f_{esc}$ can correlate with the oxygen line ratio [\ion{O}{3}]/~[\ion{O}{2}], due to a higher expected [\ion{O}{3}] flux at a given [\ion{O}{2}] flux in density bounded nebulae \citep[e.g.][]{Nakajima14}. 
\citet{Faisst16} used a compilation of eight detections and four upper limits of $f_{esc}$ to show a tentative positive correlation with [\ion{O}{3}]/[\ion{O}{2}]. Out of these sources, $Ion2$ was the sole representative of the $z>0$ universe. It is thus interesting to test whether our sources agree with this correlation. 

Of our star-forming candidates, [\ion{O}{3}]/[\ion{O}{2}] is available for GS30668 and GS14633, and they have $f_{esc}$ of $60_{-38}^{+40}\%$ and $62_{+38}^{-51}\%$ respectively. For an escape fraction of 0.6, the relationship derived in \citet[][]{Faisst16} predicts [\ion{O}{3}]/[\ion{O}{2}]$\sim11$. Interestingly, only GS30668 is close to this value, while GS14633 has a significantly lower value of [\ion{O}{3}]/[\ion{O}{2}]$\sim3$, albeit with a large uncertainty. 

It is worthwhile to turn to \citet[][]{Stasinska15} who used a large sample of galaxies with extreme [\ion{O}{3}]/[\ion{O}{2}] and a careful analysis of photoionization models to conclude that [\ion{O}{3}]/[\ion{O}{2}] on its own is an insufficient diagnostic tool for the leakage of LyC photons and must be used along with other lines like [\ion{Ar}{3}], [\ion{O}{1}] and \ion{He}{2}, and considerations of the gas covering fraction \citep[][]{Reddy16b}. It must be acknowledged that our line ratios are rather uncertain ($\sim2.5\sigma$). Followup observations of our candidates to obtain high resolution spectra, are thus important in order to make definitive statements about the [\ion{O}{3}]/[\ion{O}{2}] approach towards constraining $f_{esc}$.

\subsubsection{EW(H$\beta$)-$\beta_{UV}$}
\citet[][]{Zackrisson13} show via simulations that the EW(H$\beta$)-$\beta_{UV}$ diagram is an effective selector of high $f_{esc}$ at $z>6$, and in a follow-up study \citep[][]{Zackrisson16} conclude that a rest-frame $\mathrm{EW(H\beta)<30\angstrom}$ is sufficient to select for $f_{esc} > 0.5$ at $z\sim 7-9$. Since their conclusions and diagrams only apply to dust-free SEDs with $\beta < -2.3$ (typical of $z>6$ galaxies) we can only discuss GS30668 ($\beta = -2.23 \pm 0.08$; E(B-V)$=0$). Qualitatively, the \citet[][]{Zackrisson16} claim seems to hold, since both $Ion2$ and GS30668 have similar metallicity, age and $f_{esc}$ while occupying essentially the same point on the EW(H$\beta$)-$\beta_{UV}$ diagram (we have discussed the resemblance of $Ion2$ and GS30668 earlier in \S\ref{comparison_highz}). It will be important, however, to test such indirect methods with larger samples of directly detected LyC emitters in the future.


\section{Summary and Outlook}
In this paper we presented six galaxies which likely exhibit a large fraction of escaping ionizing photons at $z\sim2$. These are among the first sources detected in ionizing photons at significant redshift.

The novel data that made the discovery of these candidates possible came from the HDUV survey (Oesch et al., submitted), the deepest large-area UV survey undertaken by \textit{HST} to date. The F275W and F336W measurements from HDUV in combination with multi-wavelength archival GOODS+CANDELS imaging provide continuous, high resolution $HST$ photometry from the UV to near IR. 
Building on a selection method first described in \citet[][]{Vanzella15} we developed an SED-modelling  Monte-Carlo method to detect flux excesses in the UV photometry that imply LyC leakage. In our analysis we use BPASS SEDs, a newly derived dust-law for the LyC region \citep[][]{Reddy16a} and well-tested realizations of the IGM transmission \citep[][]{Inoue14}. Based on this method, we discovered six sources that have a high probability to be leaking LyC flux into the IGM and we estimate their absolute $f_{esc}$--all very high, but ranging from $>0.13$ to unity (at 90\% likelihood). A future paper in preparation will address the average escape
fraction of galaxies in the HDUV fields. In general, our findings are in
broad agreement with previous studies that found only a small fraction of
galaxies to show high $f_{esc}$. 

While our sources are clearly not representative of the average galaxy at these redshifts, we are finding evidence that the maximally achievable $f_{esc}$ is evolving with cosmic time. Currently, no source with $f_{esc}>13\%$ has been found at low redshift, while several of the individual detections at $z>2$ (including our galaxies) are consistent with $f_{esc}>50\%$ (Fig \ref{fig:summary}).

Thanks to how richly studied the GOODS fields are, we are able to draw from existing literature and ancillary data (chiefly the 3D-\textit{HST} grism survey, whose redshifts also helped in the selection) to investigate these sources in some detail. We use very deep Chandra X-ray data, Spitzer fluxes, and a Herschel study to identify three of our sources as AGN. In two of the AGN sources, it is likely that the LyC flux nevertheless predominantly originates from star-forming regions, aided by the clearing out of the ISM by the active nucleus. In the remaining three sources the ionizing radiation is likely to originate purely from stars.
 
A comparison of the star-forming galaxies in our sample with the three previously known high-$z$ LyC sources proves to be quite revealing. GS30668 and $Ion2$, extreme \ion{O}{3} emitters with two of the largest EW([\ion{O}{3}]$_{4959+5007}$ + [H$\beta$]) recorded at high-$z$ resemble each other in many aspects. In general, our candidates are $\sim L*_{UV}$ galaxies and they show relatively young stellar population ages of $\lesssim100$ Myr and little dust extinction, as has been found for previous LyC emitters.

Looking to the future, it will be important to use candidates like the ones presented here to calibrate indirect methods of estimating $f_{esc}$, since LyC photons are effectively impossible to observe beyond $z\gtrsim4.5$. This includes relationships between $f_{esc}$ and parameters like \ion{O}{3}/\ion{O}{2} and H$\beta$/$\beta_{UV}$. In this work, we show tentatively that these relations have promise. High-quality, high-resolution spectra that capture features like [\ion{O}{2}], H$\beta$, Si and Ly$\alpha$ are required for our candidates. Studies like these will allow us to infer $f_{esc}$ for galaxies directly in the epoch of reionization with James Webb Space Telescope (JWST) NIRSPEC observations in the future to derive a self-consistent picture of cosmic reionization.


\acknowledgments{ The authors thank Akio Inoue for providing Monte-Carlo realizations of the IGM transmission at high redshift.
The primary data for this work were obtained with the \textit{Hubble Space Telescope} operated by AURA, Inc. for NASA under contract NAS5-26555. Support for this work was provided by NASA through grant HST-GO-13871 from the Space Telescope Science Institute, which is operated by AURA, Inc., under NASA contract NAS 5-26555. RN thanks Yale Astronomy's Dorrit Hoffleit Undergraduate Research Scholarship, Alice \& Peter Tan, and Yale-NUS College's Summer Independent Research Program (SIRP) for their support. PO acknowledges support by the Swiss National Science Foundation.
Some of the data presented in this paper were obtained from the Mikulski Archive for Space Telescopes (MAST). }

Facilities: \facility{HST (ACS, WFC3)}.

\bibliography{MasterBiblio}
\bibliographystyle{apj}

\end{document}